\documentclass{article}

\usepackage{arxiv}

\usepackage[utf8]{inputenc} 
\usepackage[T1]{fontenc}    
\usepackage{hyperref}       
\usepackage{url}            
\usepackage{booktabs}       
\usepackage{amsfonts}       
\usepackage{nicefrac}       
\usepackage{microtype}      
\usepackage{lipsum}		
\usepackage{graphicx}
\usepackage{natbib}
\usepackage{doi}

\usepackage{graphicx}%
\usepackage{multirow}%
\usepackage{amsmath,amssymb,amsfonts}%
\usepackage{amsthm}%
\usepackage{mathrsfs}%
\usepackage[title]{appendix}%
\usepackage{xcolor}%
\usepackage{textcomp}%
\usepackage{manyfoot}%
\usepackage{booktabs}%
\usepackage{algorithm}%
\usepackage{algorithmicx}%
\usepackage{algpseudocode}%
\usepackage{listings}%
\usepackage{caption}
\usepackage{subcaption}

\title{A Finite Volume and Levenberg–Marquardt Optimization Framework for Benchmarking MHD Flows over Backward-Facing Steps}

\author{%
  \href{https://orcid.org/0009-0006-7700-7261}{\includegraphics[scale=0.06]{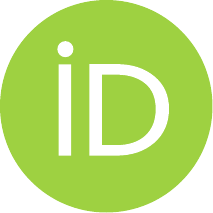}\hspace{1mm}Spyridon C.~Katsoudas}\thanks{These authors contributed equally to this work.} \\
  Department of Mathematics\\
  University of Ioannina\\
  Campus Ioannina University, Ioannina, 45110, Greece \\
  \texttt{s.katsoudas@uoi.gr} \\
  \And
  \href{https://orcid.org/0009-0006-7413-6316}{\includegraphics[scale=0.06]{orcid.pdf}\hspace{1mm}Grigorios T.~Chrimatopoulos}\footnotemark[1] \\
  Department of Mechanical Engineering\\
  University of the Peloponnese\\
  1 M. Aleksandrou Str, Koukouli, Patras, 263 34, Greece \\
  \texttt{g.chrimatopoulos@go.uop.gr} \\
  \And
  \href{https://orcid.org/0000-0001-8441-1306}{\includegraphics[scale=0.06]{orcid.pdf}\hspace{1mm}Michalis A.~Xenos} \\
  Department of Mathematics\\
  University of Ioannina\\
  Campus Ioannina University, Ioannina, 45110, Greece \\
  \texttt{mxenos@uoi.gr} \\
  \And
  \href{https://orcid.org/0000-0002-5598-564X}{\includegraphics[scale=0.06]{orcid.pdf}\hspace{1mm}Efstratios E.~Tzirtzilakis}\thanks{Corresponding author.} \\
  Department of Civil Engineering\\
  University of the Peloponnese\\
  1 M. Aleksandrou Str, Koukouli, Patras, 263 34, Greece\\
  \texttt{etzirtzilakis@go.uop.gr} \\
}

\date{}

\renewcommand{\headeright}{}
\renewcommand{\undertitle}{}
\renewcommand{\shorttitle}{A FVM and Levenberg–Marquardt Framework for MHD Flows over BFS}

\hypersetup{
pdftitle={A template for the arxiv style},
pdfsubject={q-bio.NC, q-bio.QM},
pdfauthor={David S.~Hippocampus, Elias D.~Striatum},
pdfkeywords={First keyword, Second keyword, More},
}

\begin{document}
\maketitle
\begin{abstract}
	This study examines the hydrodynamic and magnetohydrodynamic numerical solution of an electrically conducting fluid flow in a backward facing step (BFS) geometry under the influence of an external, uniform magnetic field applied at an angle. The numerical results are obtained utilizing the Finite Volume Method in a collocated grid configuration whereas the resulting system is solved directly using a Newton-like method in contrast to iterative approaches. The computed hydrodynamic results are validated with experimental and numerical studies for an expansion ratio of two. The magnetohydrodynamic case is also validated for Reynolds number $Re=380$ and Stuart number $N=0.1$ with previous numerical studies. Some applications of BFS flow under the influence of a magnetic field include metallurgical processes, cooling of nuclear reactors, plasma confinement, and biomedical applications in arteries. One of the most important findings of this study is the reduction of the reattachment point in contrast to the increase of pressure as the magnitude of the magnetic field is amplified. The magnetic field angle with the greatest influence on fluid flow has been observed to be at an angle of $\varphi = \pi/2$. In several cases, the magnetic field could substantially reduce the main flow vortex leading to a shifted reattachment point.
\end{abstract}


\keywords{ Backward Facing Step, Magnetohydrodynamics, Finite Volumes Method, Direct Numerical Solution, Point of flow reattachment }


MSC Classification: {35Q30, 65N08, 76W05}

\maketitle
\section{Introduction}\label{sec1}

The Backward-Facing Step (BFS), also referenced in the literature as “sudden expansion flow'' or “backward flow'', constitutes a classic benchmark problem in Computational Fluid Dynamics (CFD). Its significance arises from its ability to model separated flows. The flow downstream of the step point presents the separation bubble formation and reattachment processes in both the lower and upper walls. For a high Reynolds number ($Re$), multiple bubbles exist in both walls. Reattachment and separation points are highly affected by geometry, inlet-outlet boundary conditions, and heat transfer. Many numerical studies have been conducted for the investigation of the effects of the outflow boundary conditions in the BFS flow, such as Papanastasiou et al. and Gartling who highlighted the importance of such boundary conditions in the numerical solution~\cite{papanastasiou, gartling}. Examining these flows is vital due to the various real-life applications such as the flow behind buildings, airfoils at large angles of attack, or spoiler flows~\cite{spoiler,BASKARAN1996861}. Separation is a phenomenon that scientists try to avoid as it leads to flow resistance and consequently to large energy losses. 

The flow over a BFS has been established as a test problem to challenge the accuracy of numerical methods, due to the dependence of the reattachment length, $x_{r}$ on the Reynolds number and the expansion ratio (ER)~\cite{kim}. Numerous studies compare their numerical findings with experimental data, such as those of Armaly et al.~\cite{Armaly} for ER = 1.942, whereas Lee and Mateescu studied for BFS ER = 1.17 and ER = 2.0~\cite{LEE1998703}. These efforts aim to achieve accurate predictions of the reattachment point. The fundamental quantity that validates the accuracy of the method is the reattachment point of the bubble on both walls. The three-dimensional nature of the problem affects the comparison of the reattachment point in both walls, regarding two-dimensional numerical studies and three-dimensional experimental data, as Armaly et al. indicate~\cite{Armaly}. The agreement among numerical studies with experimental data is still not satisfactory and it is highlighted that the extra third dimension drastically affects the reattachment distance when $Re>500$~\cite{chen,valencia}.

As the flow separates at the corner of the step, a recirculation region forms downstream and the flow reattaches at distance $x_{r}$ from the corner step. For a large Reynolds number, additional separation and recirculation points appear in the upper wall. As $Re$ increases, multiple recirculation regions emerge in both the upper and lower walls~\cite{ERTURK2008633}. This behavior is attributed to the adverse pressure gradient associated with the expanding flow, created by a sudden change in the cross section~\cite{Thangam}.

The case of a two-dimensional, steady-state incompressible flow over a BFS has been thoroughly examined for a small Reynolds number. The main concern of scientists is the investigation of BFS flows over $Re>800$ due to the controversy on whether it is possible to obtain a stable solution. The $Re=800$ is believed to be the upper regime to obtain a stable numerical solution~\cite{ERTURK2008633}. The turbulent BFS flow has been studied meticulously by scientists; in consequence, numerical and experimental calculations have established the transitional regime to be for $Re>1150$~\cite{LEE1998703}. Numerous studies have been conducted in turbulent case as this of Erturk, Liakos and Malamataris and Choi et al. investigating the transitional regions and bifurcations that occur in two- or three-dimensional fluid flows~\cite{ERTURK2008633, LIAKOS20151, choi}. Recent studies on 3D BFS flows have examined thermal performance, focusing on how geometric changes like downstream baffles affect heat transfer. Parameters such as baffle height, thickness, and position significantly influence flow and Nusselt number distribution across varying Reynolds and Grashof numbers~\cite{Tsay2005}.

The primary distance reattachment point is a function of the expansion rate ($E=H/h$) and the Reynolds number ($Re$). In this study $Re$, is based on the height $H$, of the channel, $Re_{H}=\left(\rho u_{a} H\right)\mu$, where $u_{a}$ defined as the average value of the inlet velocity, or else the $2/3$ of $u_{max}$. In the case of ER=2.0, $Re_{H}$ is identical with, $Re_{D}=\left(\rho u_{a} D\right)\mu$, where $D$ is the hydraulic diameter defined as twice the height of the inlet region upstream of the geometry ($D=2h$). In the scope of this study, the expansion rate is selected at 2.0 because it is the most studied case in the literature and facilitates comparison. As the expansion rate increases, the reattachment point decreases as Erturk presents in~\cite{ERTURK2008633}. The geometry of BFS has a dimensionless length of $L=15.0$, height of $H=1.0$, the step height is $h=0.5$. Consequently, the expansion ratio is equal to 2.0 and the step location on the $x$-axis is at $x=0.675$ or normalized as $x=1.35h$, where $h$ is the step height. Demuren demonstrated a study and showed that the numerical solution remained unaffected by the channel length when the ratio, $L/H > 7$ which stands in the current study~\cite{demuren}. 

Despite its relative geometric simplicity, the BFS problem encapsulates complex dynamics that challenge both experimental and CFD approaches, especially when it extends to magnetohydrodynamic (MHD) contexts. A uniform magnetic field reduces the fluid momentum~\cite{davidson2017introduction}. This effect is due to the Lorentz force in the momentum equations that opposes fluid flow. Although the application of such magnetic field has been studied in various scenarios, such as in the lid driven cavity, aneurysms, and stenosis, the application of a magnetic field in a BFS problem is limited to our knowledge~\cite{gurbuz2015mhd, CHERKAOUI2023107281, TZIRTZILAKIS200866}. Its implementation in the BFS flow remains scarcely explored, highlighting the novelty of the present study~\cite{ABBASSI2007231, gurbuz2015mhd}. Although the numerical methodology used in the aforementioned studies differs significantly compared with the present one, the numerical results are in very good agreement. Magnetohydrodynamics (MHD) has been an active area of research among scientists and engineers due to its wide range of applications in industrial and bioengineering research areas~\cite{reviewofmhd, RASHIDI2017358}.

\begin{figure}
    \centering
    \includegraphics[scale=0.4]{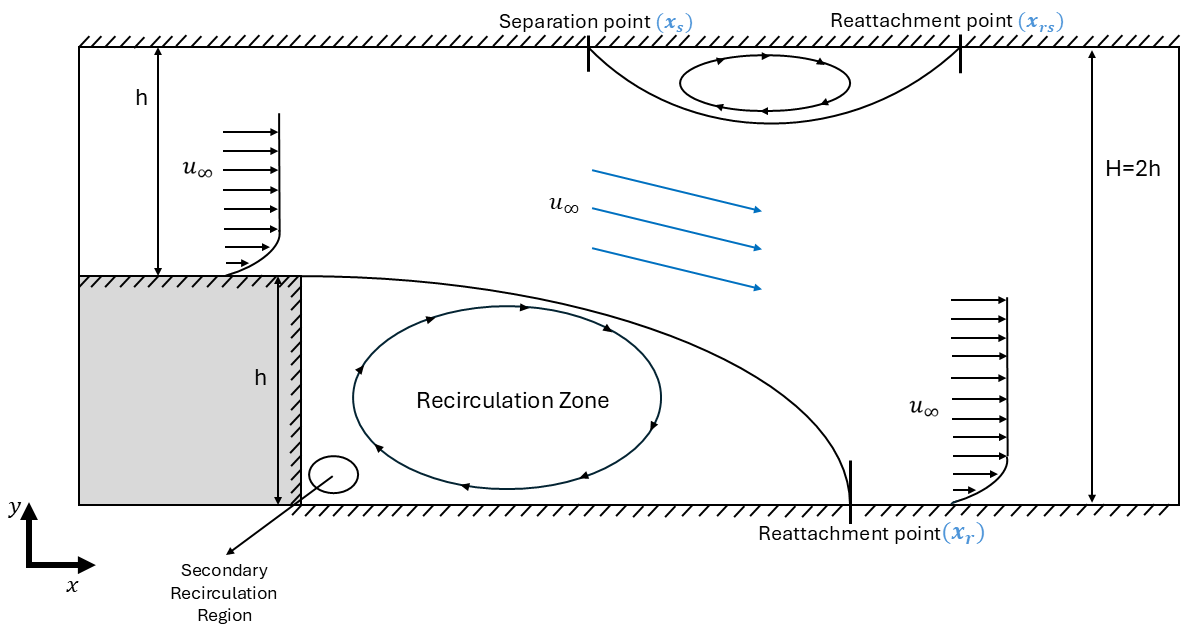}
    \caption{The geometry of BFS with the recirculation zone, the secondary recirculation region where a second smaller vortex exists and the potential flow where the velocity is equal to $u_{\infty}.$} 
    \label{Geometry_backstep}
\end{figure}

Figure~\ref{Geometry_backstep}, illustrates the geometry of the BFS in the present study. A recirculation zone is observed to the right of the BFS accompanied by the formation of vortices. Boundary layers are observed in the upper region of the step and in the region after the reattachment point.

The novelty of this study arises from the extension of MHD flow over a BFS in various angles of magnetic field and different Stuart numbers. Additionally, the numerical procedure of FVM with the Levenberg-Marquardt algorithm is validated in the MHD and hydrodynamics cases. The results are in excellent agreement with the calculations of the normalized reattachment point ($x_{r}/h$), the length of the upper wall bubble $(x_{rs}-x_{s})/h$ and the velocity profile in distance $x/h=6$ and $x/h=14$. We have developed an advanced numerical algorithm based on Newton-like methods, such as the Levenberg-Marquardt approach for the solution of a strongly nonlinear system of partial differential equations (PDEs). This approach provides robust and reliable results and always converges to the solution. A numerical code was developed in Matlab (MathWorks, Inc.). So, the literature gap is addressed with this study, presenting robust results on MHD effect on the flow field in different angle configurations, highlighting the pressure field in all the cases studied. \\

\section{Mathematical Formulation}

\subsection{Magnetic field configuration}
\begin{figure}
    \centering
    \includegraphics[scale=2.0]{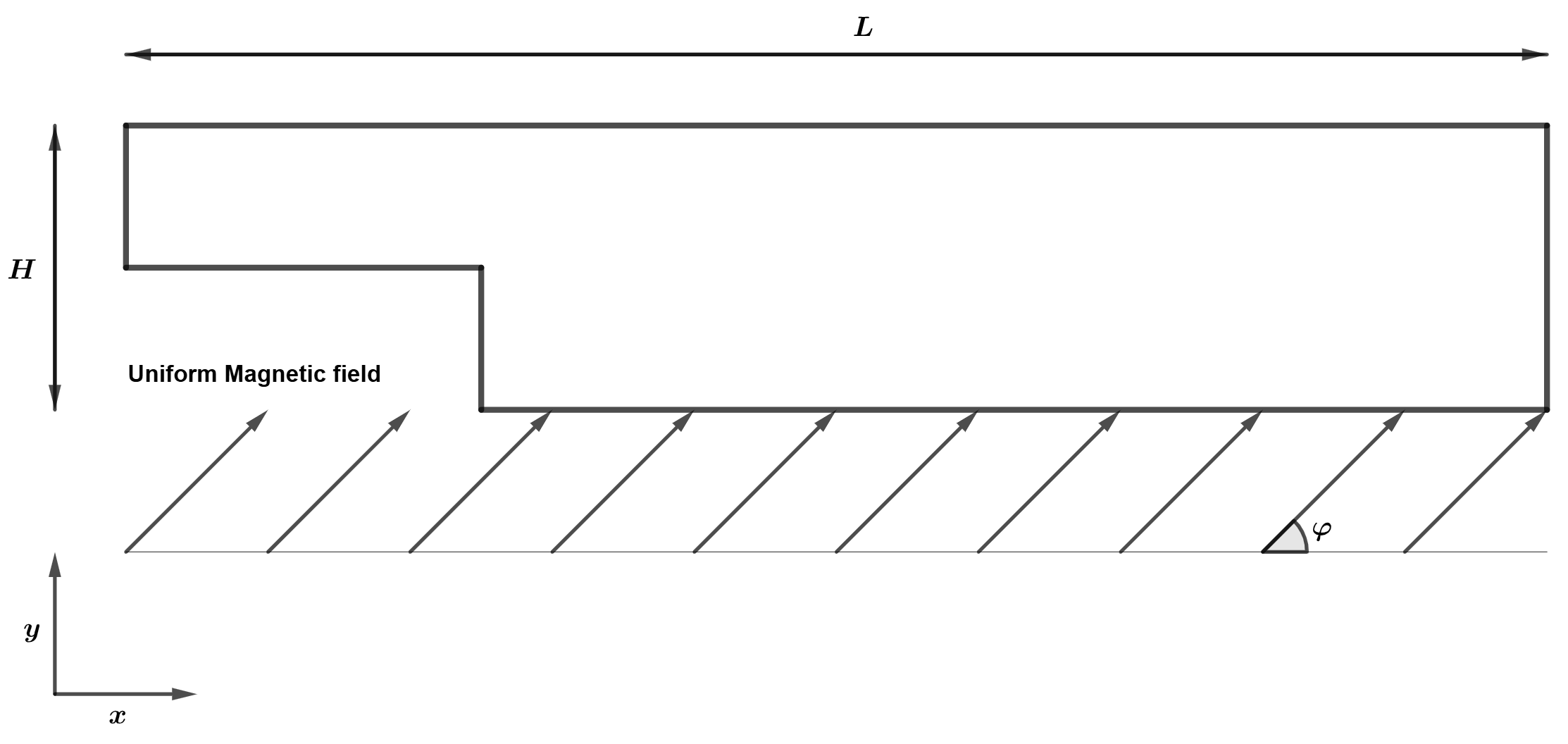}
    \caption{Backward-facing step for a uniform magnetic field of an angle $\varphi < \pi/2$.}
    \label{Backstep_MHD_FHD}
\end{figure}

In this study, a uniform magnetic field is applied to the fluid flow at an angle, $\varphi$. This type of magnetic field is applied to the entire fluid. In mathematical form, the uniform magnetic field is given by the magnetic induction~\cite{raptis2014finite},
\begin{equation}\label{MHDB0}
    \bar{B'}=\left(B'_{x'},B'_{y'}\right)=\left(B_{0}cos\varphi,B_{0}sin\varphi\right)=B_{0}\left(cos\varphi,sin\varphi\right)
\end{equation}
where $B'_{x'}$ and $B'_{y'}$ are the magnetic induction components (uniform magnetic field) and $B_{0}$ is the magnetic field magnitude measured in Tesla ($T$),
\begin{equation}
    B'=\sqrt{{B'}^{2}_{x'}+{B'}^{2}_{y'}}=B_{0}\left(cos^{2}\varphi+sin^{2}\varphi\right)=B_{0}.    
\end{equation}
The influence of the magnetic field on fluid flow is maximized when the magnetic field is applied vertically, $\varphi = \pi/2$. The change in angle below $\varphi=\pi/2$ results in a minimized effect of the magnetic field, as shown in Figure~\ref{Backstep_MHD_FHD}. The numerical results for different angles are presented in a later section.

\subsection{Dimensional and Non-dimensional governing equations}
The governing equations that formulate the flow of an electrically conducting fluid under the influence of an externally applied, uniform magnetic field are the Navier-Stokes along with the Conservation of Mass equations~\cite{raptis2014finite},
\begin{equation}\label{xmomenondim}
    \rho\left(u'\frac{\partial u'}{\partial x'}+v'\frac{\partial u'}{\partial y'}\right)=-\frac{\partial p'}{\partial x'}+\mu\left(\frac{\partial^2 u'}{\partial x'^2}+\frac{\partial^2 u'}{\partial y'^2}\right)-\sigma\left(B'^{2}_{y'}u'-B'_{x'}B'_{y'}v'\right),
\end{equation}
\begin{equation}\label{ymomenondim}
    \rho\left(u'\frac{\partial v'}{\partial x'}+v'\frac{\partial v'}{\partial y'}\right)=-\frac{\partial p'}{\partial y'}+\mu\left(\frac{\partial^2 v'}{\partial x'^2}+\frac{\partial^2 v'}{\partial y'^2}\right)-\sigma\left(B'^{2}_{x'}v'-B'_{x'}B'_{y'}u'\right),
\end{equation}
\begin{equation}\label{consnondim}
    \frac{\partial u'}{\partial x'}+\frac{\partial v'}{\partial y'}=0,
\end{equation}
where $u'$ and $v'$ are the dimensional fluid velocity components, $p'$ is the dimensional fluid pressure. $\rho$ is the fluid density, measured in kilograms per cubic meter ($kg/m^2$) $\mu$ is the fluid viscosity, measured in Pascal-seconds ($Pa \cdot s$). $\sigma$ is the electrical conductivity, measured in Siemens per meter ($S/m$). The induced magnetic field is considered negligible~\cite{rajput2011rotation}.

To obtain the non-dimensional forms of equations \eqref{xmomenondim}-\eqref{consnondim}, the following non-dimensional quantities are introduced,
\begin{equation}\label{nondimnums}
    x=\frac{x'}{H}, \quad y=\frac{y'}{H}, \quad u=\frac{u'}{u_{a}}, \quad v=\frac{v'}{u_{a}}, \quad p=\frac{p'}{\rho u_{a}^{2}}, \quad B_{x}=\frac{B'_{x'}}{B_{0}}, \quad B_{y}=\frac{B'_{y'}}{B_{0}}.
\end{equation}
where $H$ is the channel height measured in meters ($m$), $u_{a}$ measured in meters per second ($m/s$). The quantity $B_{0}$ is the magnetic field magnitude measured in Tesla ($T$),
\begin{equation}
    \bar{B}=\left(B_{x},B_{y}\right)=\left(\cos\varphi,\sin\varphi\right).
\end{equation}
\vskip 0.1 cm
The non-dimensional governing equations are written in closed form as,
\begin{equation} \label{xmome}
    \frac{\partial u^2}{\partial x}+\frac{\partial uv}{\partial y}=-\frac{\partial p}{\partial x}+\frac{1}{Re}\left(\frac{\partial^2 u}{\partial x^2}+\frac{\partial^2 u}{\partial y^2}\right)-N\left(B^{2}_{y}u-B_{x}B_{y}v\right),
\end{equation}
\begin{equation} \label{ymome}
    \frac{\partial uv}{\partial x}+\frac{\partial v^2}{\partial y}=-\frac{\partial p}{\partial y}+\frac{1}{Re}\left(\frac{\partial^2 v}{\partial x^2}+\frac{\partial^2 v}{\partial y^2}\right)-N\left(B^{2}_{x}v-B_{x}B_{y}u\right),
\end{equation}

where $Re_{H}$ is the Reynolds number and $N$ is the Stuart number given by \cite{raptis2014finite},
\begin{equation}\label{non-dim-nums}
    Re_{H}=\frac{\rho \: u_{a}\: H}{\mu}, \quad N=\frac{\sigma H B_{0}^{2}}{\rho u_{a}}.
\end{equation}


\subsection{Boundary conditions}
\noindent The boundary conditions for the problem under consideration are,
\begin{equation}\label{bc}
    \begin{gathered}
        u\left(y\right)= 
            \begin{cases}
            \quad \quad\quad \quad \quad 0,& y \in[0,0.5)\\
            u_{max}(-16y^2+24y-8),  & y\in[0.5,1] \\
                
            \end{cases},
        \quad v=0, \quad \text{at the channel inlet}, \\
        u=0,\quad v=0, \quad \text{at the top and bottom walls}, \\
        p=0, \quad \text{at the channel outlet.}
    \end{gathered}
\end{equation}
The inflow velocity profile is identical to the studies we compare with. The value of average velocity $u_{a}$ is equal to $1.0$ and is used for the calculation of $Re_{H}$. At the outlet channel region of BFS, Neumann boundary conditions are applied for the two components of velocity and zero pressure Dirichlet boundary condition is applied. On the walls of the BFS geometry, no-slip boundary condition is enforced for the fluid velocity. The case of $ER = 2.0$ is fundamental and the most studied, therefore the choice for the expansion rate to compare with as many studies as possible.

\section{Numerical methodology}
The discretization approach used in the current study is the Finite Volume Method (FVM), a second-order accuracy method, due to the central scheme used to discretize the governing equations~\cite{versteeg2007introduction}. This method transforms the system of nonlinear partial differential equations into a nonlinear system of algebraic equations over each finite volume. The computational domain is divided into a finite number of volumes, composed of partitions of the domain. By integrating the governing equations~\eqref{xmome}-\eqref{ymome} over the control volume, the required system is obtained, where in a residual form is provided by, 
\begin{equation}\label{xmomefvm}
    \begin{gathered}
        f_{1}=\frac{1}{2}\Delta y\left(u_{E}^{2}-u_{W}^{2}\right)+\frac{1}{2}\Delta x\left(u_{N}v_{N}-u_{S}v_{S}\right)+\Delta y\left(p_{E}-p_{P}\right)-\\
        \frac{1}{Re}\left(\frac{\Delta y}{\Delta x}\left(u_{E}-2u_{P}+u_{W}\right)+\frac{\Delta x}{\Delta y}\left(u_{N}-2u_{P}+u_{S}\right)\right)+\\  N\left(cos^{2}\left(\varphi\right)u_{P}-sin\left(\varphi\right)cos\left(\varphi\right)v_{P}\right),
    \end{gathered}
\end{equation}
\begin{equation}\label{ymomefvm}
    \begin{gathered}
        f_{2}=\frac{1}{2}\Delta y\left(u_{E}v_{E}-u_{W}v_{W}\right)+\frac{1}{2}\Delta x\left(v_{N}^{2}-v_{S}^{2}\right)+\Delta x\left(p_{E}-p_{P}\right)-\\
        \frac{1}{Re}\left(\frac{\Delta y}{\Delta x}\left(v_{E}-2v_{P}+v_{W}\right)+\frac{\Delta x}{\Delta y}\left(v_{N}-2v_{P}+v_{S}\right)\right)+\\  N\left(sin^{2}\left(\varphi\right)v_{P}-sin\left(\varphi\right)cos\left(\varphi\right)u_{P}\right),
    \end{gathered} 
\end{equation}
\begin{equation}\label{consfvm}
    \begin{gathered}
        f_{3}=\frac{1}{2}\left(u_{E}-u_{W}\right)+\frac{1}{2}\Delta x\left(v_{N}-v_{S}\right).
    \end{gathered}
\end{equation}
$\Delta x$ and $\Delta y$ are the dimensions of the control volume and $E,W,P,S$ and $N$ are the East, West, Centroid, South and the North of the control volume.

The numerical solution is obtained by solving the equations~\eqref{xmomefvm}-\eqref{consfvm} using a Newton-like method~\cite{quarteroni}. In general, Newton-like methods use the Jacobian matrix, calculated by,
\begin{equation}
    J = \frac{\partial \left(f_{1},f_{2},f_{3}\right)}{\partial \left(u,v,p\right)}
\end{equation}
The functions $f_{1},f_{2}$ and $f_{3}$ are the residual equations obtained by the discretization scheme equations~\eqref{xmomefvm}-\eqref{consfvm} and the variables $\Tilde{u}, \Tilde{v}$ and $\Tilde{p}$ are the vectors of the unknown parameters of the fluid velocities and pressure. In large-scale problems, such as the one presented in this study, the evaluation of the Jacobian matrix, $J$, can be computationally expensive~\cite{Davis_Rajamanickam_Sid-Lakhdar_2016}. The sparsity of the Jacobian matrix was utilized, saving substantial amount of memory while deceasing the computational time. It has been observed that the number of zero elements increases as the grid becomes finer, implying that the number of elements equal to zero is vastly greater than the non-zero ones~\cite{Davis_Rajamanickam_Sid-Lakhdar_2016}.  

Solving the algebraic system with Newton-like methods provides a different approach compared to other solvers such as SIMPLE-PISO and Stream-Function Vorticity formulation. The proposed method does not use additional equations to simplify the nonlinear system. The coupled system is solved simultaneously at each Newton iteration, while the Jacobian matrix is calculated once per iteration, saving computational time and space. Convergence is obtained when the equations~\eqref{xmomefvm}-\eqref{consfvm} are as close to zero as possible. A small tolerance value ($\varepsilon=10^{-10}$) is defined to stop the solver procedure. If this criterion is satisfied, a vector solution is obtained which contains the resolved unknown parameters $u,v$ and $p$~\cite{quarteroni}.

\section{Results and Discussion}
In this work, we study the BFS in the hydrodynamic and magnetohydrodynamic case. The BFS is a valuable test problem for testing new developed codes and the accuracy of new numerical methods. The flow over BFS combines the phenomenon of separation, recirculation region, and the existence of a primary vortex generated in the recirculation zone. In some special cases with certain expansion rates and a specific Reynolds number $Re$, a secondary vortex emerges at the low corner of the step. The application of the MHD principles results in the disturbance of fluid flow. The momentum is reduced along the vortex length, and the reattachment point is shifted. The area after the reattachment point, defined as the region in which the boundary layer is recovered. 

The numerical results presented are obtained through a mesh grid of the size of $199\times109$ which results in control volumes providing a higher spatial resolution and ensuring greater accuracy in capturing gradients and flow characteristics. During the study of the BFS flow a grid independence study has been conducted using finer grids. Maximum differences of $2.3\%$ have been observed for the reattachment point at the bottom wall for $Re$ numbers of $100$ to $800$. Thus, the coarser grid was used to obtain the numerical results. 

\subsection{Hydrodynamic Case}
Backward facing step flow in the hydrodynamic case is studied for  different Reynolds numbers, $Re$ from $100$ to $800$ with step $100$. The results include the $u$-velocity component, pressure distribution, skin friction coefficient, $C_{f}$ and vital values such as these of normalized reattachment point, $x_{r}/h$ and normalized length of the bubble in the upper wall, $(x_{rs}-x_{s})/h$, to evaluate the results with other studies.

\begin{figure}
    \centering
    \includegraphics[scale=0.3]{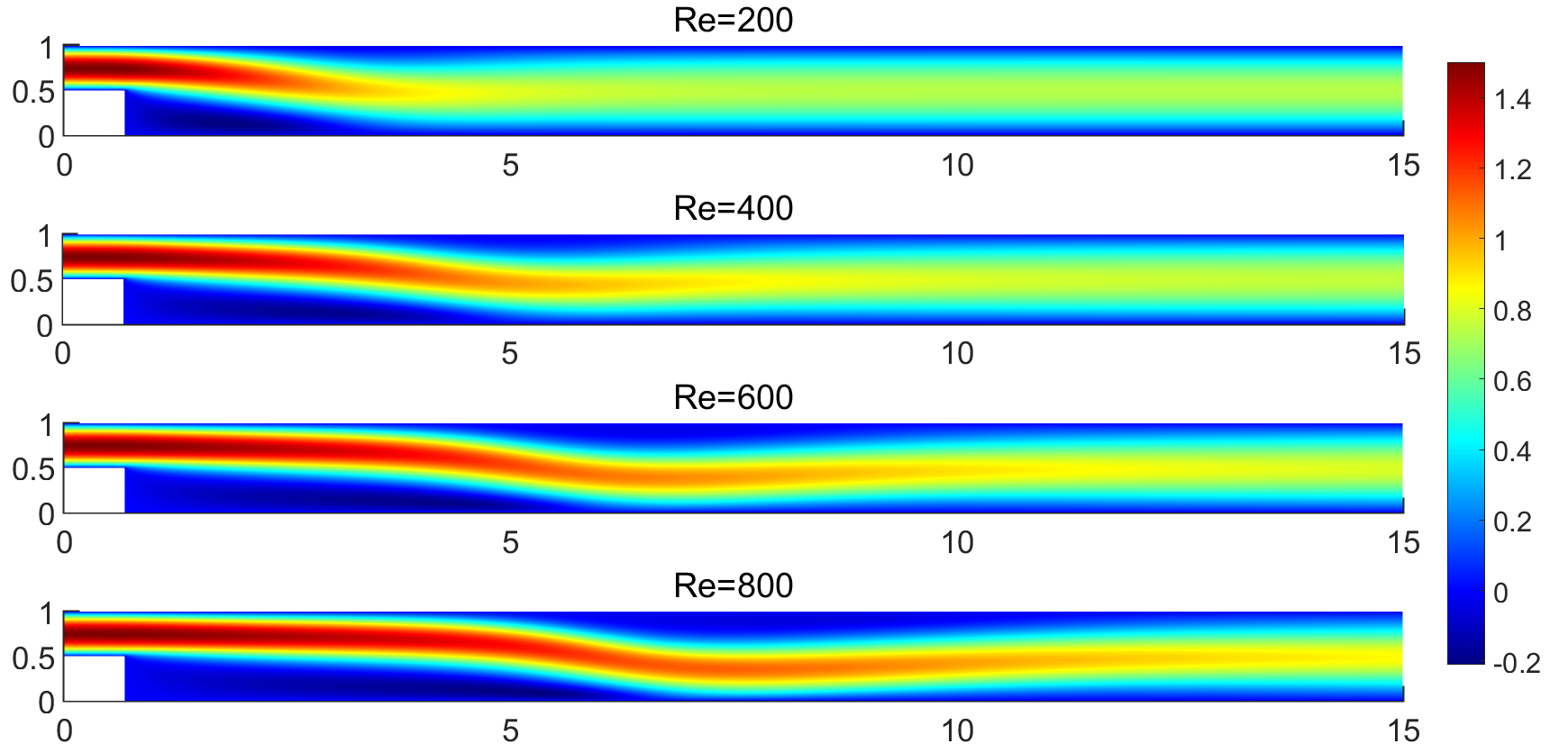}
    \caption{$u$-velocity contours for $Re=200, \ 400, \ 600$ and $800$, respectively.}
    \label{uvel}
\end{figure}

\begin{figure}
    \centering
    \includegraphics[scale=0.3]{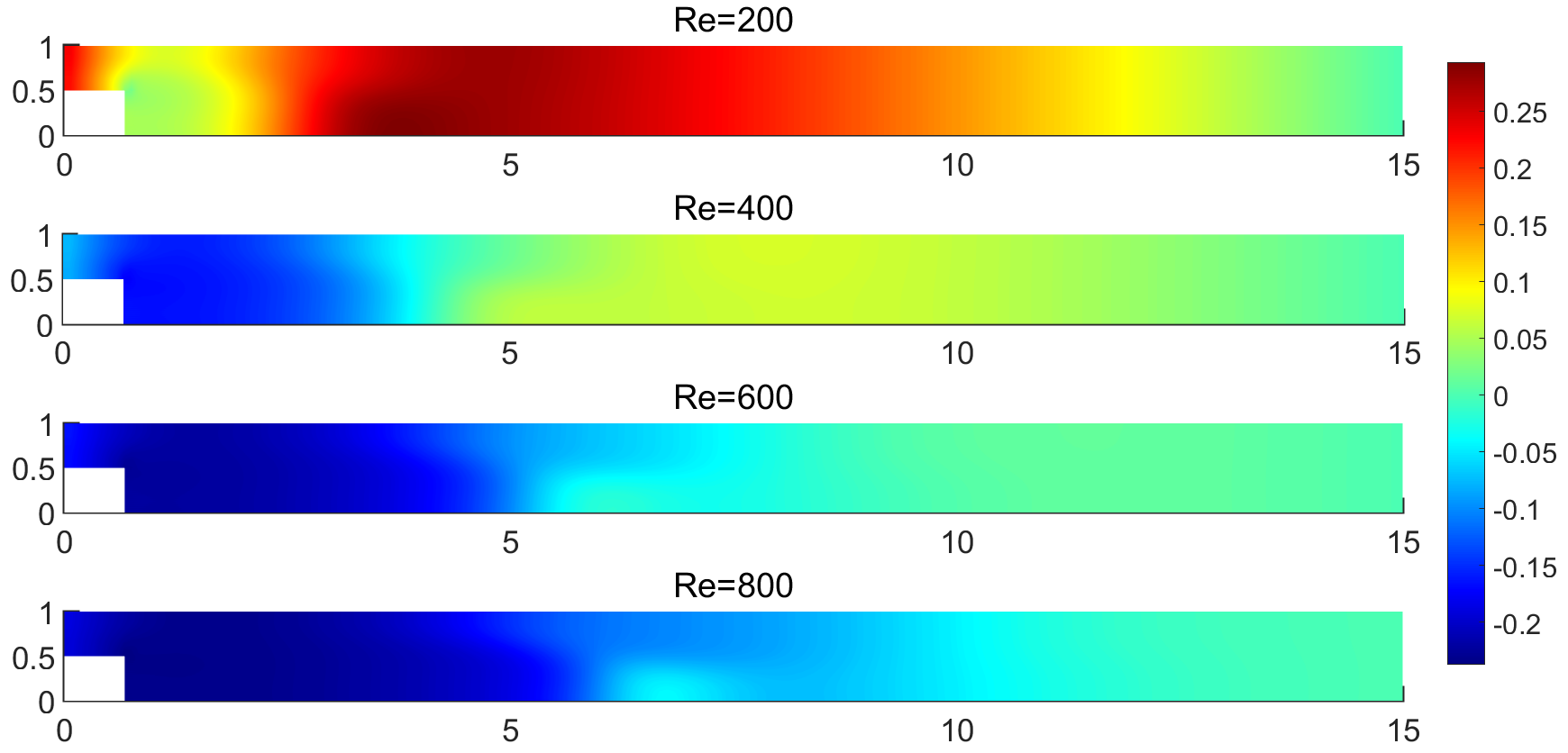}
    \caption{Pressure distribution contours for $Re=200, \ 400, \ 600$ and $800$, respectively.}
    \label{press}
\end{figure}

The $u$-velocity can be observed in Figure~\ref{uvel} for four different cases $Re=200, \ 400, \ 600$ and $800$ with maximum inlet-velocity of $u_{max}=1.5$. In the upstream region of the BFS, the flow is fully developed and remains symmetrical. After the step, separation of flow causes a significant change in the velocity profiles, creating a recirculation zone. The profile in this region shows negative velocities near the wall due to flow reversal. Furthermore, the pressure distribution for the aforementioned cases is presented in Figure~\ref{press}. As $Re$ increases, the fluid flow presents more intense recirculation and a longer reattachment length. In consequence, the primary vortex gets wider. This occurs because the inertia of the flow dominates the viscous forces as $Re$ increases. A second bubble is emerging in the upper wall at $Re=400$. This is anticipated and validates the results of the developed methodology~\cite{Armaly, ERTURK2008633}. 

\noindent
The streamlines highlight the presence of the primary vortex defining the recirculation region. The reattachment point is calculated by the stresses on the wall, $\tau_{w}$ or the skin friction coefficient, $C_{f}$. Mathematically, $\tau_{w}$ and $C_{f}$ are linked by the expression, $C_{f}=\tau_{w}/\left(\rho u_{o}^2\right)$. Consequently, the reattachment point is defined as the point that $\tau_{w}$ and $C_{f}$ are becoming zero. Similarly, the separation point and the second reattachment point on the upper wall are defined. Table~\ref{table1}, shows the normalized distances of the reattachment and separation points for various $Re$. In the last column, the length of the upper-wall bubble is computed to facilitate comparison with other studies.

\begin{table}[h]
\caption{Normalized reattachment distances of lower wall $x_{r}/h$, upper wall $x_{rs}/h$ separation point, distance in upper wall $x_{s}/h$, and length of bubble in upper wall, $(x_{rs}-x_s)/h$}
\label{tab1}
\begin{tabular*}{\textwidth}{@{\extracolsep{\fill}}ccccc}
\toprule
\textbf{$Re$} & \textbf{$x_r/h$} & \textbf{$x_s/h$} & \textbf{$x_{rs}/h$} & \textbf{$(x_{rs}-x_s)/h$} \\
\midrule
100 & 3.00  & --    & --    & --    \\
200 & 4.95  & --    & --    & --    \\
300 & 6.75  & --    & --    & --    \\
400 & 8.10  & 9.00  & 11.55 & 2.55  \\
500 & 9.30  & 9.30  & 14.70 & 5.40  \\
600 & 10.05 & 9.90  & 17.25 & 7.35  \\
700 & 10.95 & 10.35 & 19.65 & 9.30  \\
800 & 11.55 & 10.80 & 21.90 & 11.10 \\
\hline
\end{tabular*}
\label{table1}
\footnotetext{Note: ``--'' indicates values not applicable for those Reynolds numbers.}
\end{table}

\begin{figure}
    \centering
    \includegraphics[scale=0.38]{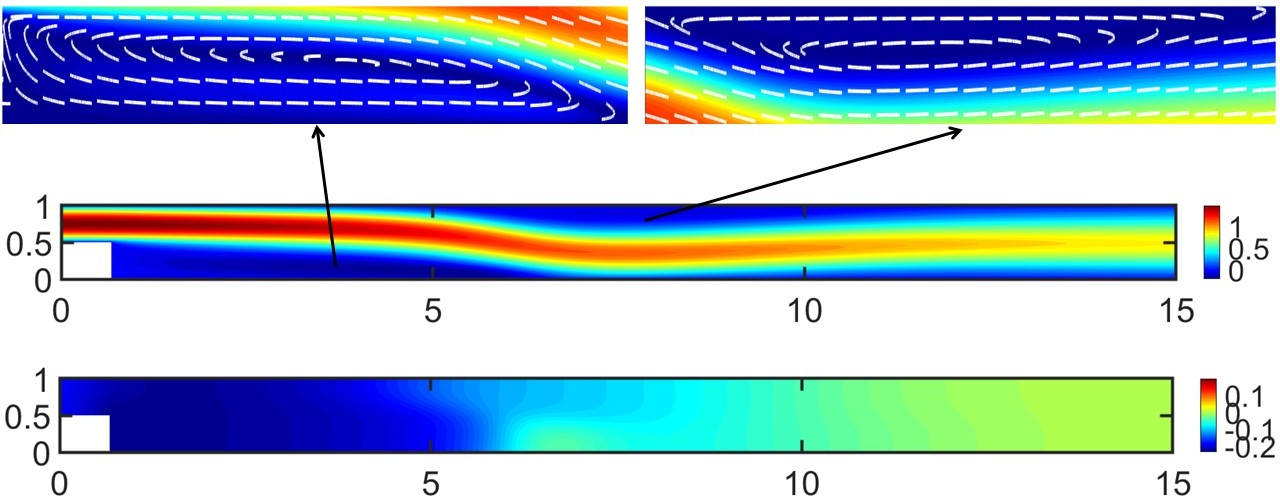}
    \caption{BFS for $u$-velocity zoomed with streamlines in the lower bubble (left) and upper bubble (right). Pressure distribution in the second figure. Figures correspond to $Re = 800$.}
    \label{block}
\end{figure}

The formation and length of the two bubbles have become the most important tests for validation in BFS flows. In Figure~\ref{block} the two bubbles for $Re=800$ are presented with streamlines. In the pressure distribution, the step in geometry lead to flow separation. The pressure in the region upstream of the step, shows a typical pressure distribution for a channel. After the step, there is a sudden pressure drop due to flow separation, which is recovering progressively as the flow approaches the reattachment point.  After this point, the pressure gradually reaches the boundary condition value applied at the end of the geometry. 

\begin{figure}
    \centering
    \includegraphics[scale=0.17]{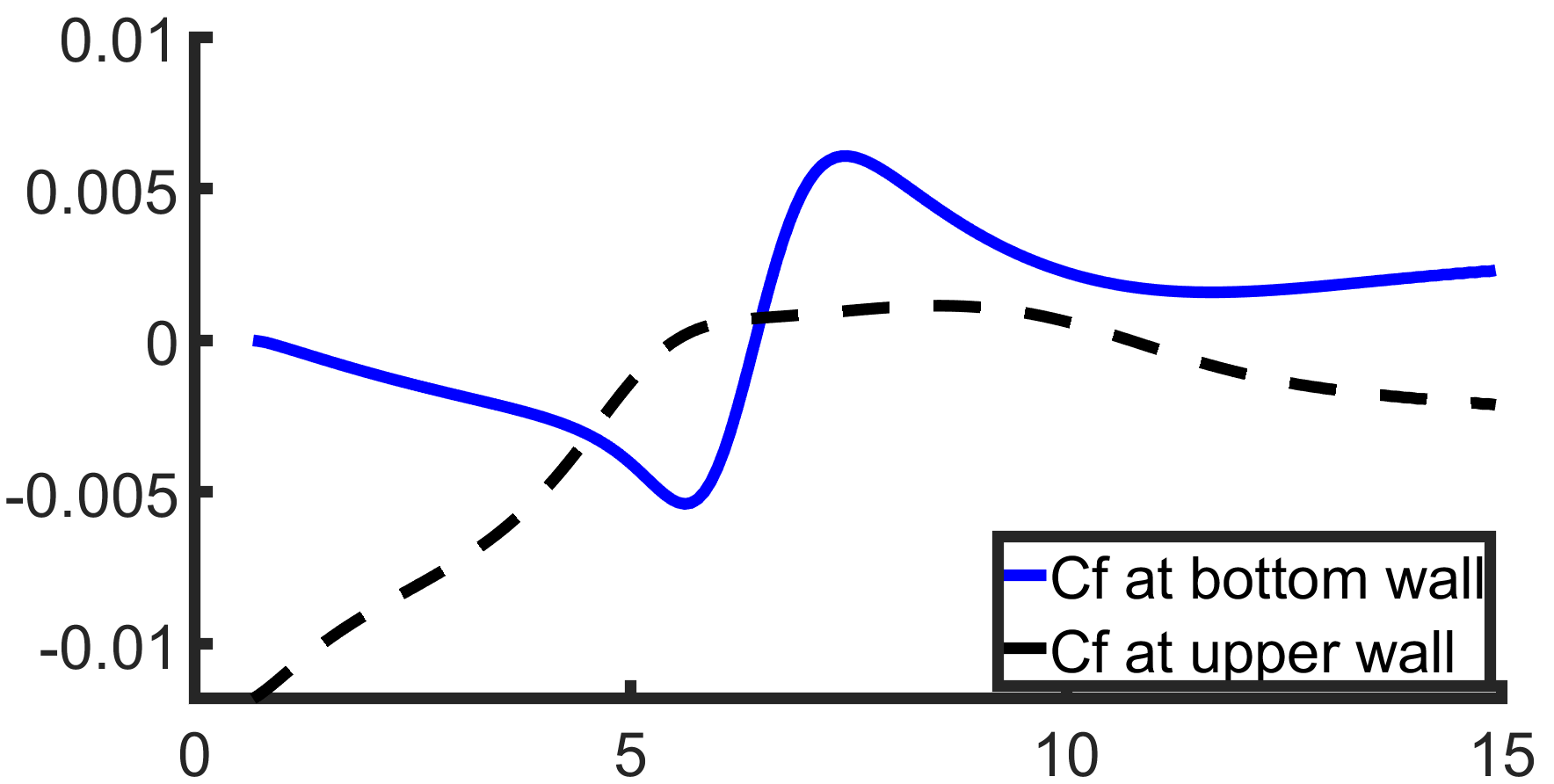}
        \caption{Skin friction coefficient in both upper (solid) and lower (dashed) wall.}
    \label{Cf}
\end{figure}

In Figure~\ref{Cf}, the skin friction coefficient on both walls is presented, after the step. It can be observed that on lower wall, $C_{f}$ is equal to zero in one location, at the reattachment point, $C_{f}$ on the upper wall is equal to zero in two locations. The first one defines the separation or else detachment point, $x_{s}$ and the second one defines the reattachment point of the upper bubble, $x_{rs}$. 

\subsection{Validation}

Validation of the results is essential for further investigation of the MHD effects in the flow over a BFS. A key indication of the robustness of the findings is the agreement between the reattachment point on the lower wall and the length of the upper wall bubble. The upper bubble exists for $Re$ greater than $400$. Numerous studies have verified $Re = 400$ as the critical value for the appearance of the upper bubble.

In Table~\ref{tab2} the reattachment point is compared to several other studies that compare the normalized reattachment point and achieve very good agreement for the numerical results in the case of $Re=800$. Additionally, the normalized length of the upper bubble presents very good agreement in comparison with other authors. In these two critical values, the small differences can be explained by the difference of numerical schemes and discretization methods. The comparison of the results with other works is evaluated as very good, considering that most authors use the stream function-vorticity formulation, which differs drastically from this methodology and results in small differences.

\begin{table}[h]
\centering
\caption{Comparison of $x_{r}/h$ and $(x_{rs}-x_{s})/h$ values with percentage error.}
\label{tab2}
\begin{tabular}{@{}lcccc@{}}
\toprule
\textbf{Authors} & \textbf{$x_{r}/h$} & \textbf{\% Error in $x_{r}/h$} & \textbf{$(x_{rs}-x_{s})/h$} & \textbf{\% Error in $(x_{rs}-x_s)/h$} \\
\midrule
Rogers \& Kwak~\cite{Rogers}              & 11.48  & 0.61\% & 11.07  & 0.27\% \\
Barton~\cite{barton}                      & 11.51  & 0.35\% & 11.52  & 3.78\% \\
Erturk~\cite{ERTURK2008633}               & 11.834 & 2.46\% & 11.077 & 0.21\% \\
Kim \& Moin~\cite{kim}                    & 11.90  & 3.03\% & 11.50  & 3.60\% \\
Lee \& Mateescu~\cite{LEE1998703}         & 12.00  & 3.90\% & 11.00  & 0.90\% \\
Guj \& Stella~\cite{guj}                  & 12.05  & 4.33\% & 10.60  & 4.50\% \\
Grigoriev \& Dargush~\cite{Grigoriev}     & 12.18  & 5.45\% & 11.24  & 1.26\% \\
Keskar \& Lyn~\cite{Keskar}               & 12.19  & 5.54\% & 11.25  & 1.35\% \\
Gartling~\cite{gartling}                  & 12.20  & 5.63\% & 11.26  & 1.44\% \\
Gresho et al.~\cite{gresho}               & 12.20  & 5.63\% & 11.26  & 1.44\% \\
Present study  
& 11.55  & --     & 11.10  & --     \\
\hline
\end{tabular}
\footnotetext{Note: ``--'' indicates base reference values from the present study, hence no error calculated.}
\end{table}

\begin{figure}[H]
    \centering
    \includegraphics[scale=0.3]{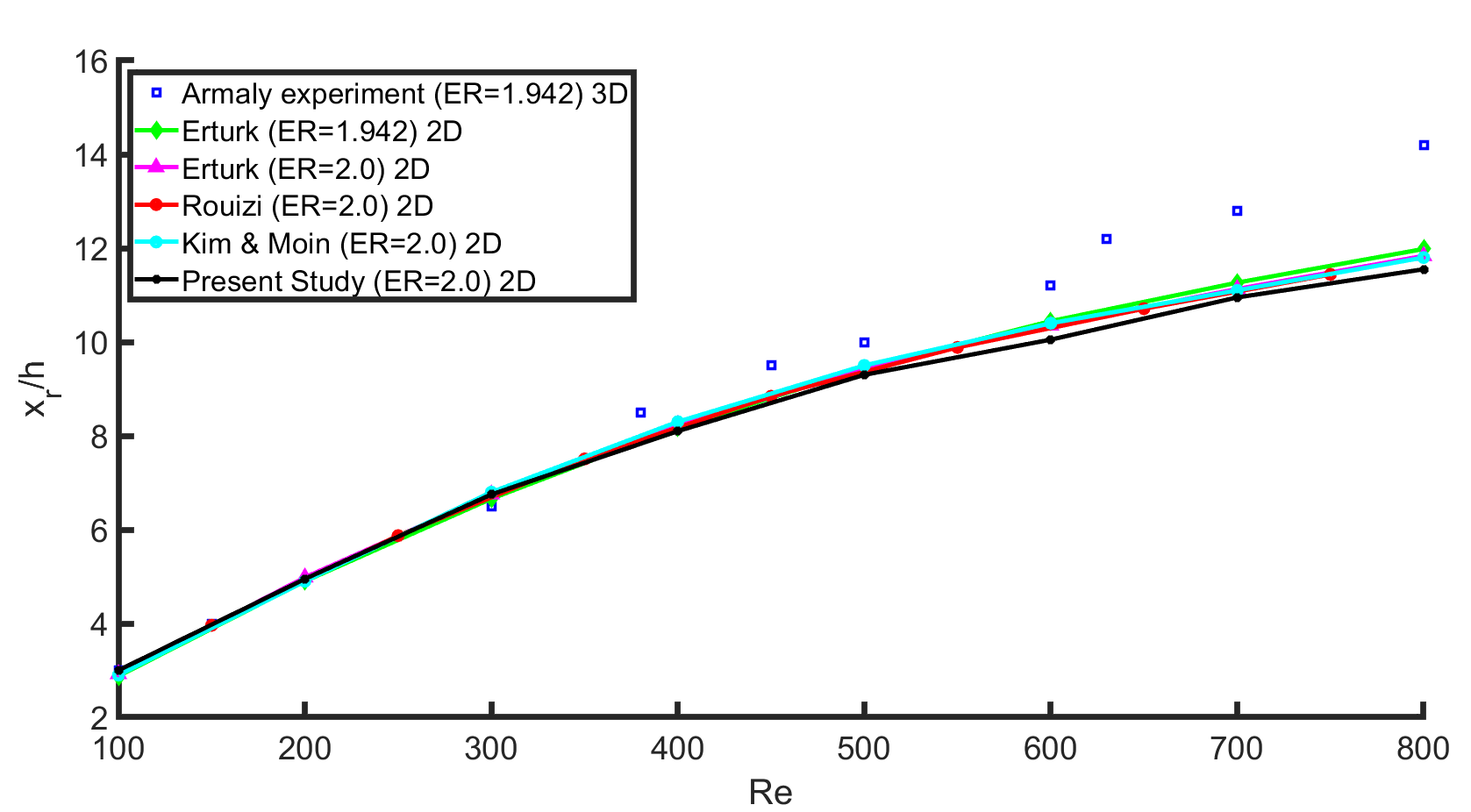}
    \caption{Comparison of present study normalized reattachment point ($x_{r}/h$) with experimental and numerical data~\cite{Armaly, ERTURK2008633, kim, rouizi}.}
    \label{comp}
\end{figure}

As Armaly et al. have demonstrated, the experimental studies vary enough from the numerical ones, at the range of $Re > 400$ due to the three-dimensionality of BFS flow, causing the two-dimensional numerical results to differ moderately from the experimental measurements~\cite{Armaly}. As depicted in Figure~\ref{comp} experimental measurements of Armaly, begin to differ noticeably after $Re = 400$, as the numerical results of several studies tend to follow an underestimate of the experimental results for an expansion ratio of $1.942$. Numerical studies of the most influential works in flow over BFS, are compared with our results and present very good agreement for all ranges of Reynolds numbers from $100$ to $800$. 

\begin{figure}
\centering
  \begin{subfigure}[b]{0.475\textwidth}
    \includegraphics[width=0.6\textwidth]{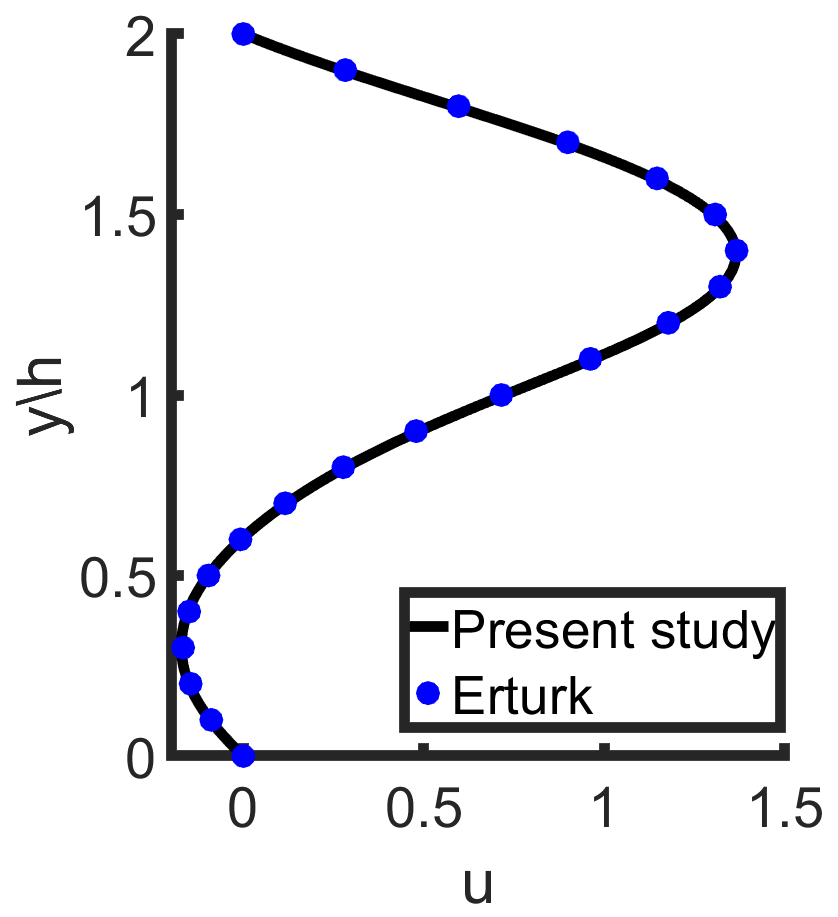}
    \caption{$x/h=6$}
    \label{xh6}
  \end{subfigure}
  \hfill
  \begin{subfigure}[b]{0.475\textwidth}
    \includegraphics[width=0.6\textwidth]{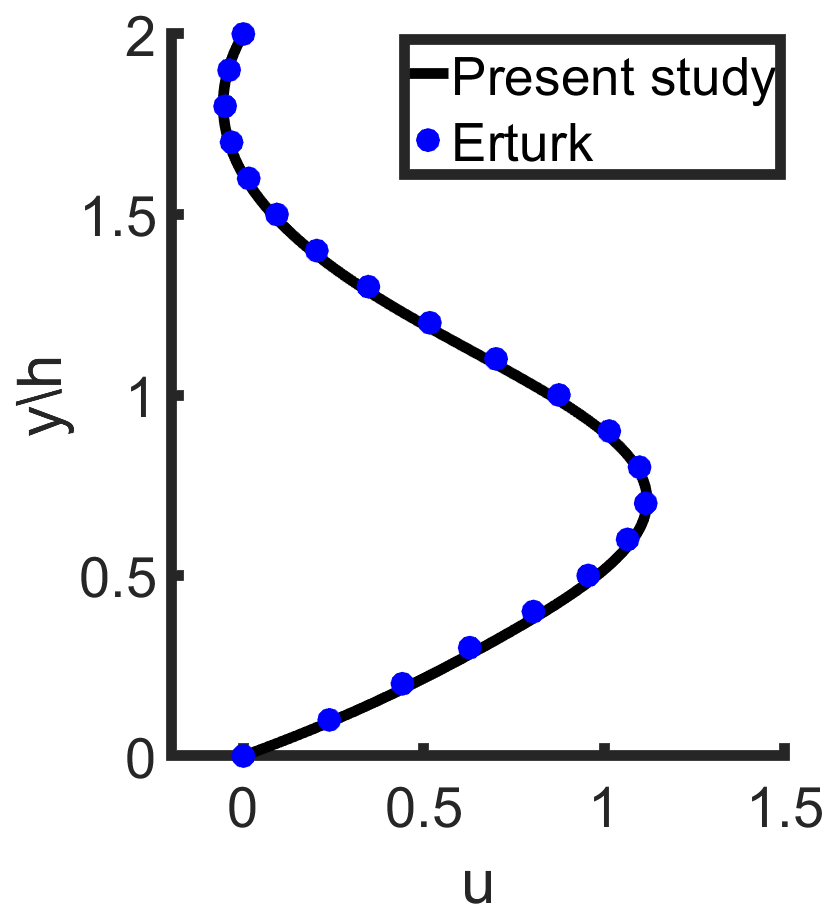}
    \caption{$x/h=14$}
    \label{xh14}
  \end{subfigure}
  \caption{Validation of horizontal profile in $x/h=6$ and $x/h=14$ with a comparison of the present study (black solid line) with Erturk profile (blue dotted line) \cite{ERTURK2008633}.}
  \label{hydro1}
\end{figure}

The negligible differences presented in the implemented comparison can be explained in the different discretization scheme, the different solver for the nonlinear system of equations, or in the slightly shorter length section we use compared to other studies. Another crucial aspect is the $u$-velocity profile. In the present study, the velocity profile at the downstream location of the step at $x/h = 6$ and $x/h = 14$ is compared to the corresponding result of Erturk~\cite{ERTURK2008633}, presenting a perfect fit, Figure~\ref{hydro1}. 

Finally, our results are satisfactory as the validation of both lower- and upper-wall bubbles are in acceptable range. This outcome provides a solid foundation for proceeding with the implementation of the fluid flow over BFS adding the MHD effects, presented in the next section.

\section{Magnetohydrodynamic flow validation}
In the present study, a uniform magnetic field is applied in various magnitudes and angles. The hydrodynamic case revealed the creation of vortices and the numerical results were validated with published studies. The application of a uniform magnetic field, due to Lorentz forces, opposes the flow. The application of the uniform magnetic field is examined in two cases, (a) the application of different magnetic field magnitudes, while the magnetic field is vertically applied to the lower wall, (b) the application of a constant magnetic field applied in various angles. A comparison with Abbassi and Nassrallah has a small error of $u$-velocity distribution at $x/h$ validating the MHD approach.

\begin{figure}[H]
\centering
  \begin{subfigure}[b]{0.475\textwidth}
    \includegraphics[width=0.6\textwidth]{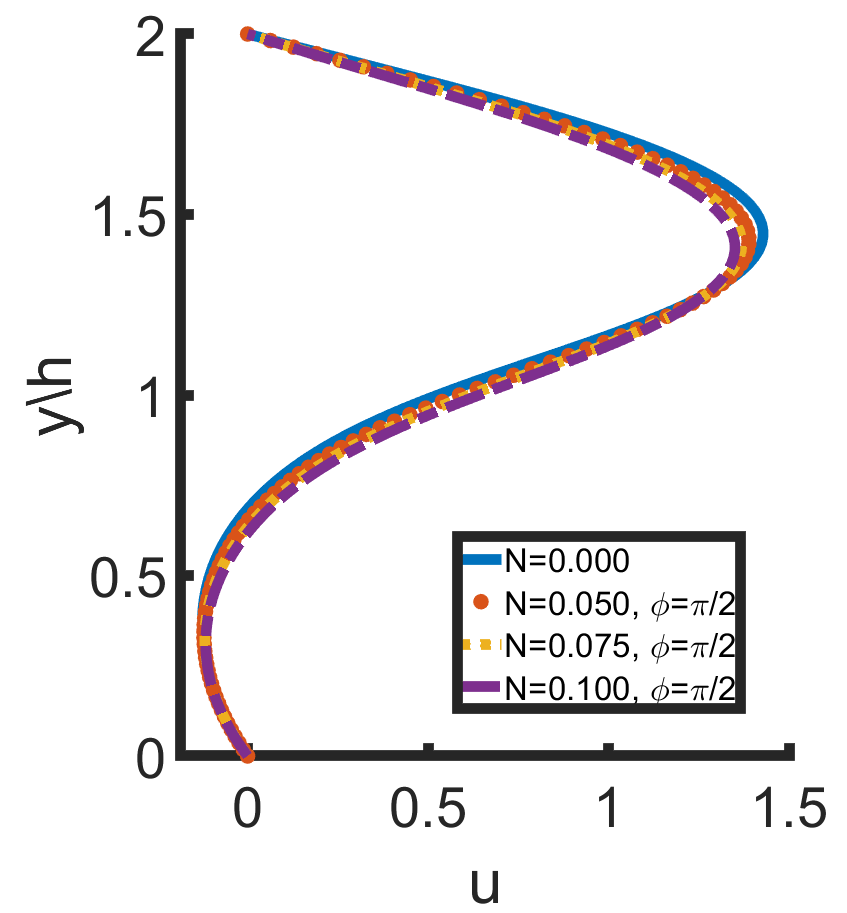}
    \caption{$x/h=2.1$}
    \label{mhdx2100}
  \end{subfigure}
  \hfill
  \begin{subfigure}[b]{0.475\textwidth}
    \includegraphics[width=0.6\textwidth]{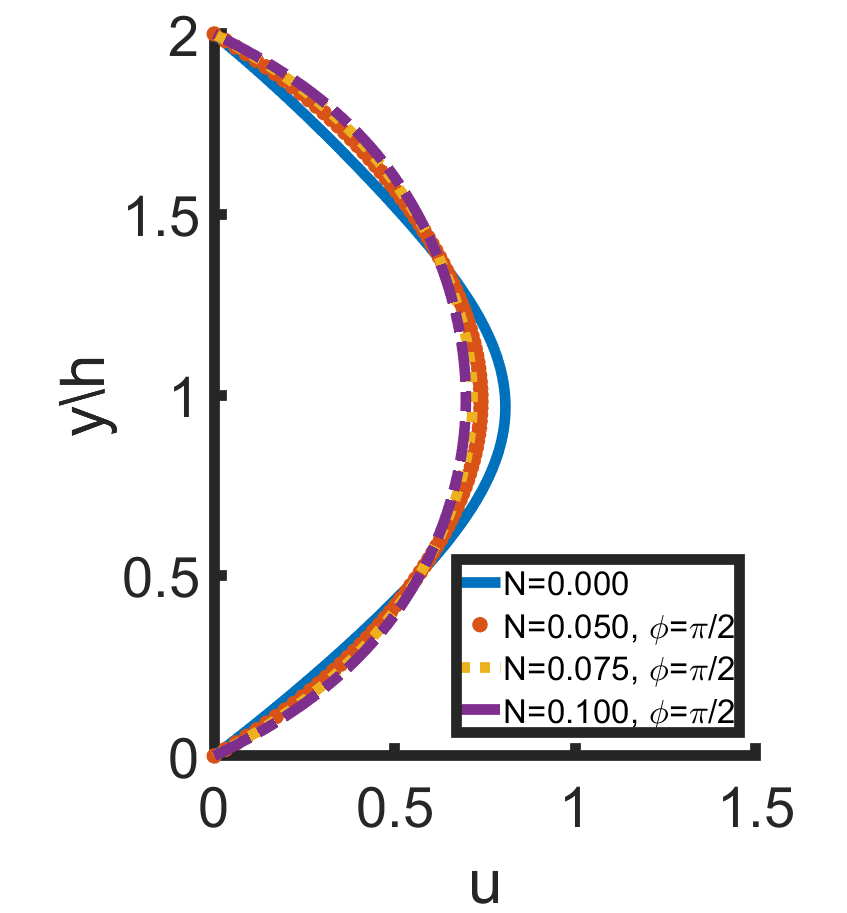}
    \caption{$x/h=20.01$}
    \label{mhdx2001}
  \end{subfigure}
  \caption{Validation of horizontal profile in $x/h=2.1$ and $x/h=20.1$ for various Stuart numbers for $Re=300$.}
\end{figure}

\begin{figure}[H]
    \centering
    \includegraphics[scale=0.2]{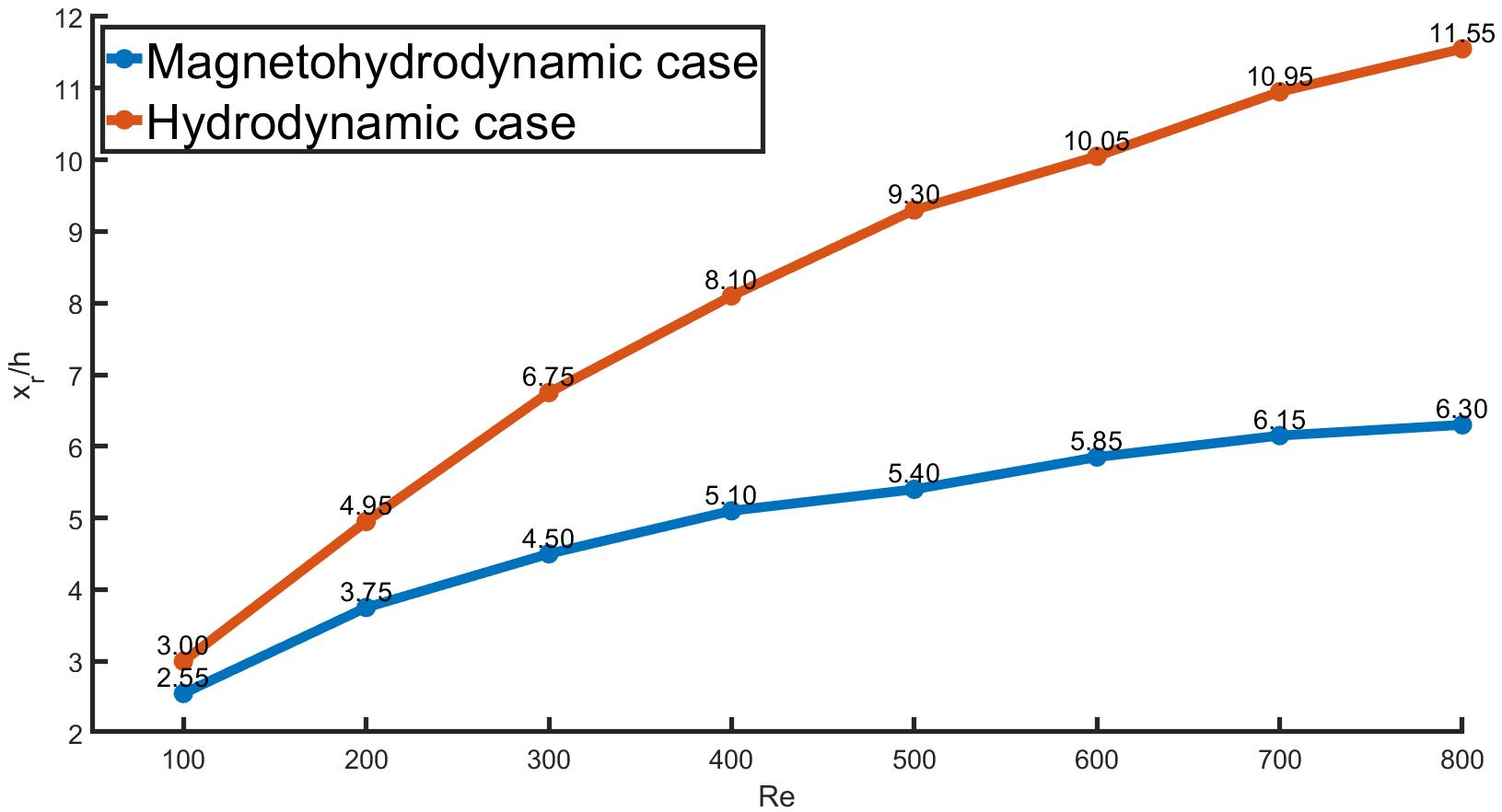}
    \caption{Comparison of the normalized reattachment points for the Hydrodynamic (orange) and Magnetohydrodynamic (blue) case, respectively, for $N = 0.1$ and $\varphi = \pi/2$.}
    \label{comparisonMHDHYD}
\end{figure}

Figure~\ref{mhdx2100} and~\ref{mhdx2001} shows the $u$-velocity at $x/h=2.1$ and $x/h=20.01$ inside the channel, respectively. At the first point the velocity presents negative values due to the creation of the vortex, whereas at the second point the velocity is closer to a parabolic profile, indicating that the flow develops. The results are also in good quantitative agreement with those of study~\cite{gurbuz2015mhd}.
Figure~\ref{comparisonMHDHYD} shows the change in the normalized reattachment point $x_{r}/h$ for the hydrodynamic and magnetohydrodynamic case, respectively. Due to the effect of the magnetic field, the reattachment point decreases. It should be noted that the vortices that are present for $Re > 400$, in the hydrodynamic case at the upper wall are completely vanished, since $C_{f}$ does not change sign. The reattachment point in the case of $Re = 800$, $N = 0.1$ and $\varphi = \pi/2$ is close to that of $Re = 300$ in the hydrodynamic case, indicating that there is no vortex in the upper wall. 

\subsection{Vertically applied magnetic field}
\begin{figure}[H]
    \centering
    \includegraphics[scale=0.3]{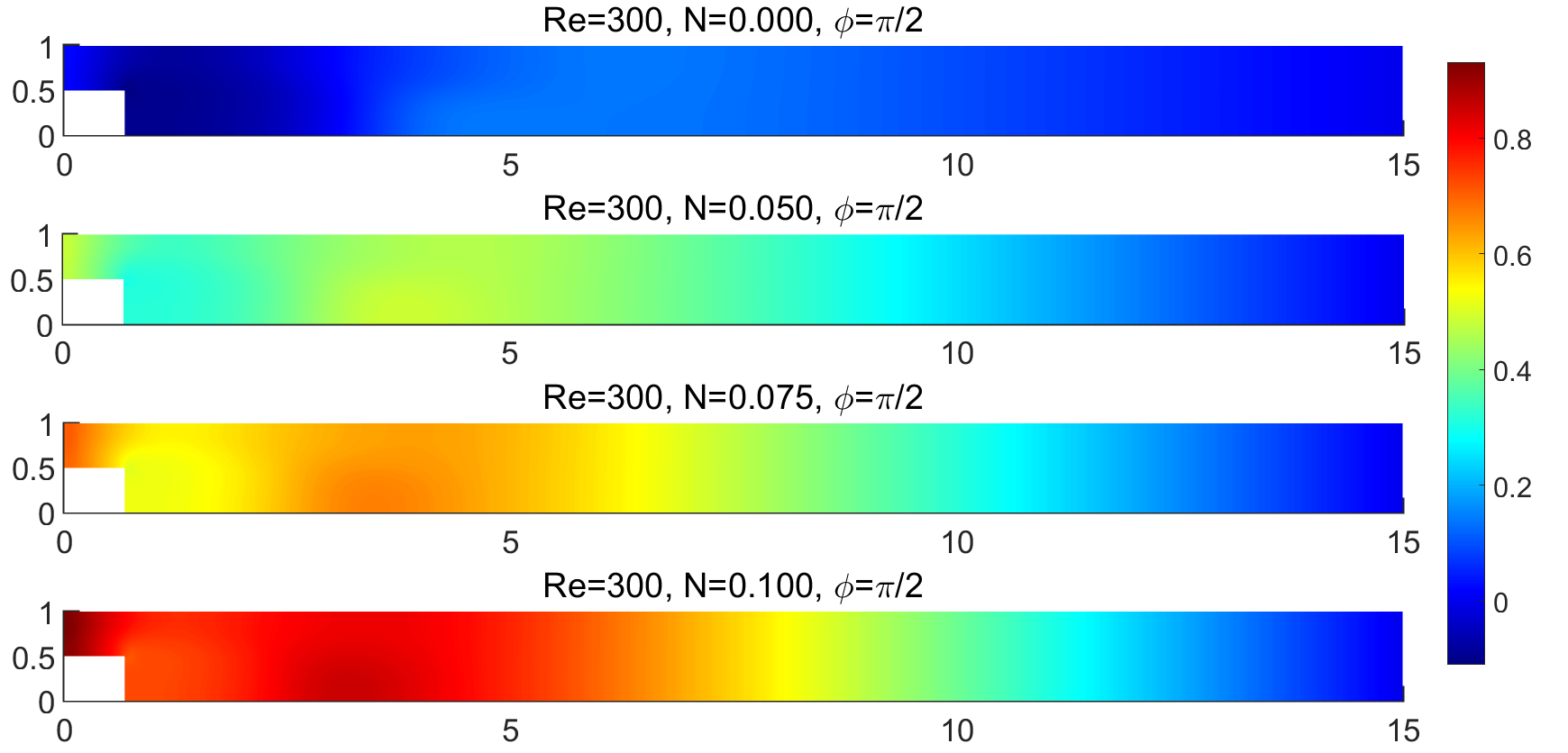}
    \caption{Pressure contours for $Re = 300$, $\varphi = \pi/2$ and Stuart number $N = 0.0$ (hydrodynamic case), $N = 0.050, 0.075$ and $0.1$, respectively.}
    \label{p_magnetic_fields}
\end{figure}

\begin{figure}[H]
    \centering
    \includegraphics[scale=0.45]{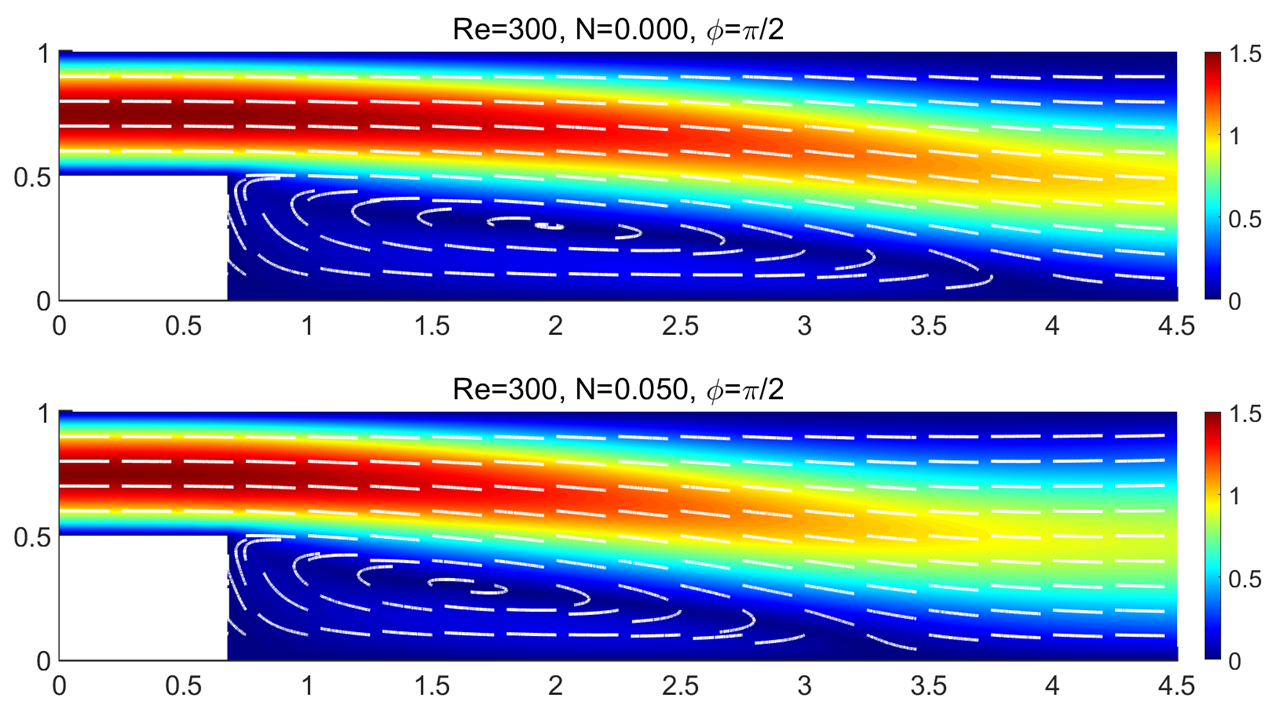}  \includegraphics[scale=0.45]{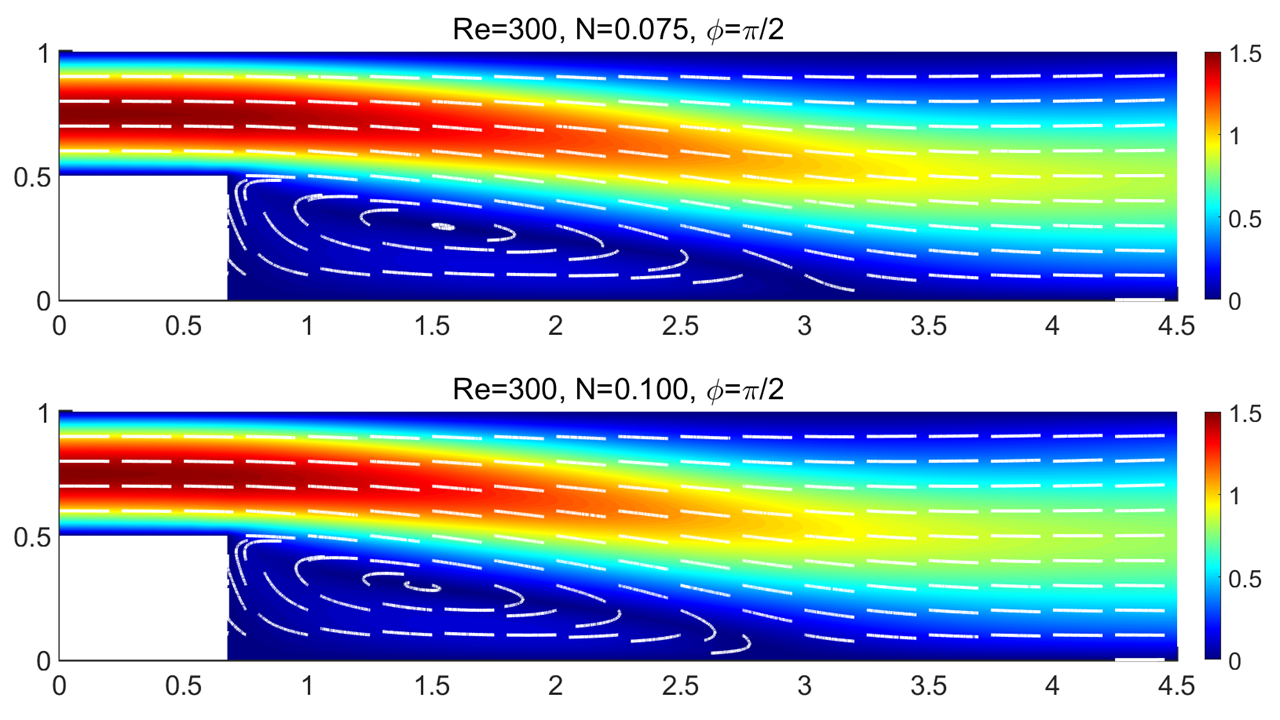}
    \caption{Creation of vortices for $Re = 300$, $\varphi = \pi/2$ and Stuart number $N = 0.0$ (hydrodynamic case), $N = 0.050, 0.075$ and $0.1$, respectively.}
    \label{vortex_magnets}
\end{figure}

The numerical simulations showed that the velocity decreases as the magnetic field magnitude increases. The pressure increases as the magnetic field magnitude is amplified, since the magnetic field opposes the fluid. A visual representation for the pressure increase is shown in Figure~\ref{p_magnetic_fields}. As a result of the velocity drop, the length of the vortex is reduced as the magnitude of the magnetic field increases, Figure~\ref{vortex_magnets}. A visual representation of the reduction of the vortex in each case is shown in Figure~\ref{vortex_magnets}, as the magnetic field increases.

\subsection{Magnetic field magnitude with various angles}
\begin{figure}[H]
    \centering
    \includegraphics[scale=0.45]{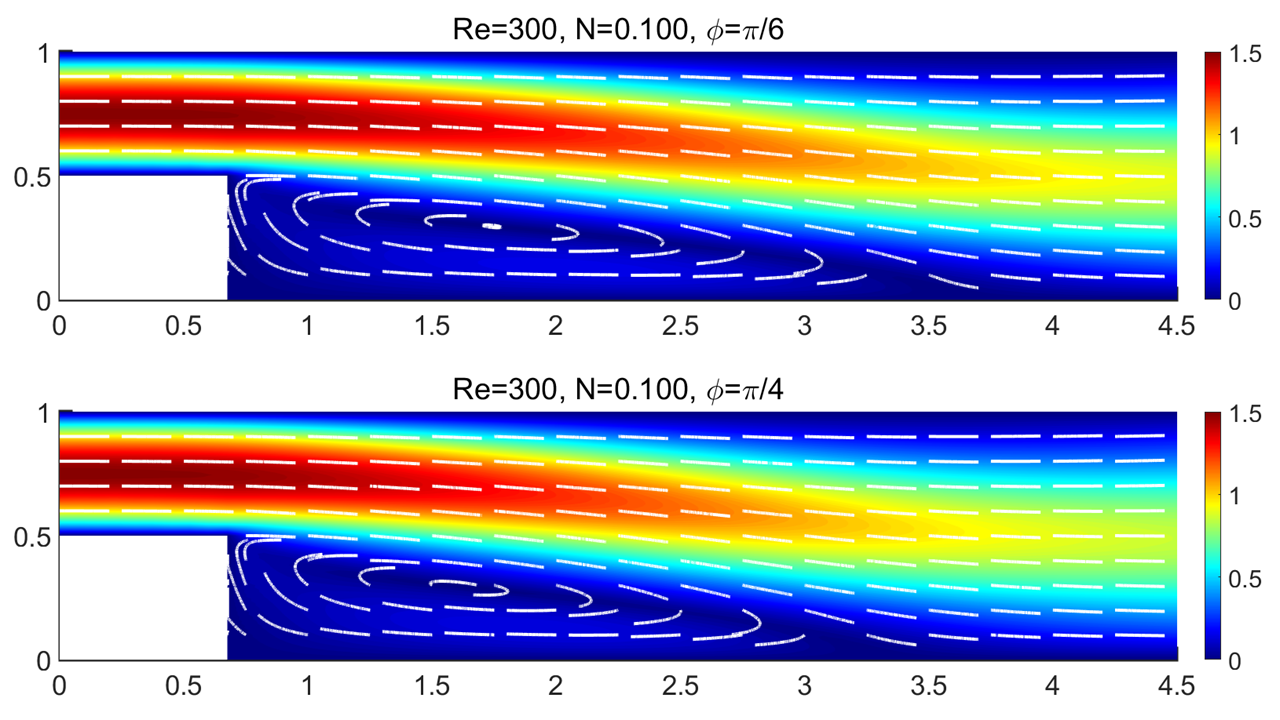}  \includegraphics[scale=0.45]{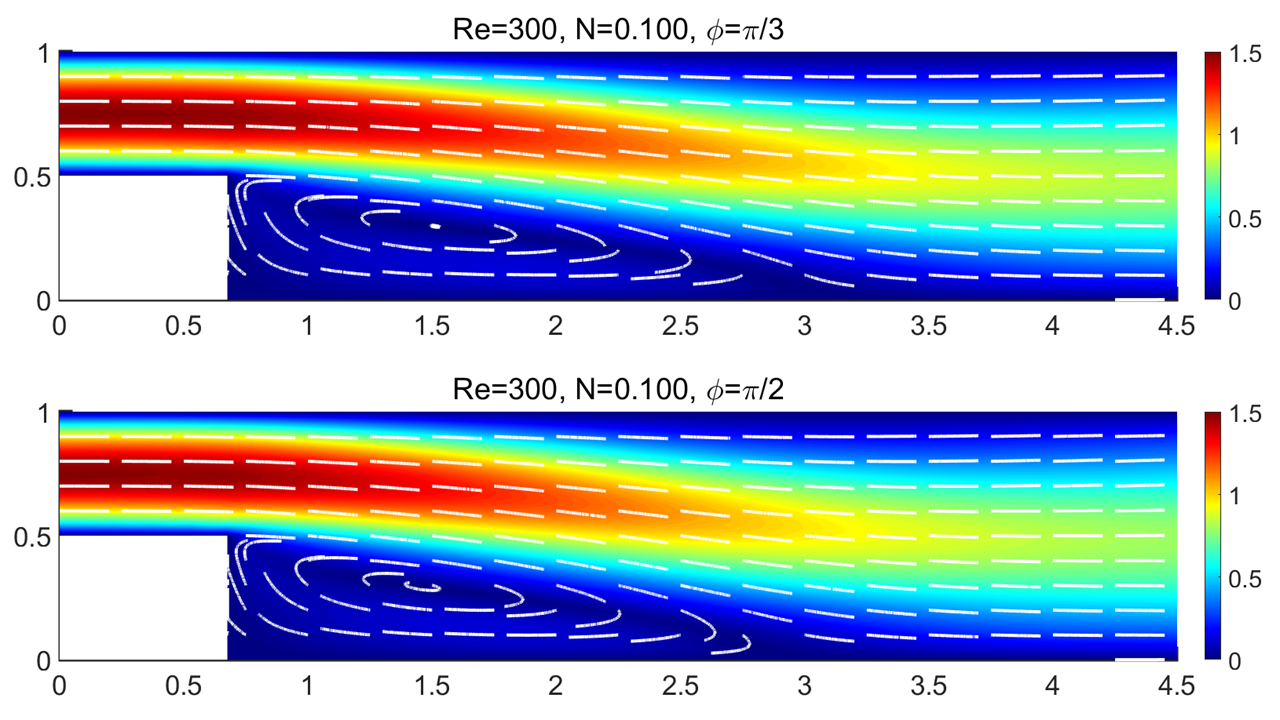}
    \caption{Creation of vortices for $Re=300$, $N=0.1$ and angles $\varphi=\pi/6, \pi/4, \pi/3$ and $\pi/2$, respectively.}
    \label{vortex_angles}
\end{figure}

The $u$-velocity is depicted for $Re=300$ and $N=0.1$, for different angles. It can be seen that the case of $\varphi=\pi/6$, is closer to the hydrodynamic case compared to the case of $\varphi = \pi/2$ where the fluid flow is considerably reduced. The vortex length becomes smaller, as the angle increases, Figure~\ref{vortex_angles}. 

\begin{figure}[H]
    \centering
    \includegraphics[scale=0.3]{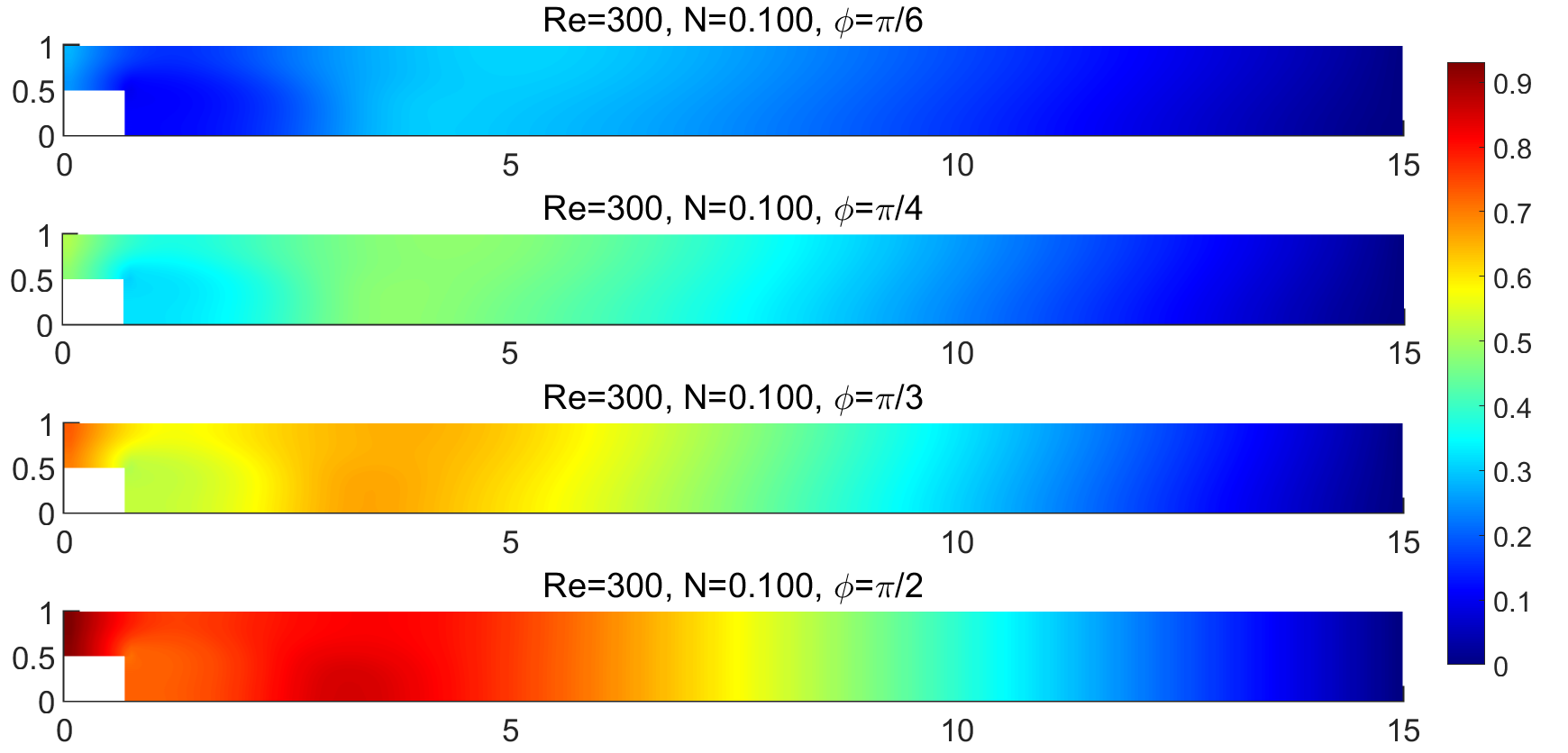}
    \caption{Pressure contours for $Re=300$, $N=0.1$ and $\varphi=\pi/6, \pi/4, \pi/3$ and $\pi/2$, respectively.}
    \label{p_angles}
\end{figure}

In this case, the Stuart number was $N = 0.1$, whereas the angle was set to $\varphi = \pi/2, \pi/3, \pi/4$ and $\pi/6$, respectively. The effect of the magnetic field on the pressure is visible at Figure~\ref{p_angles}. Lateral pressure is observed in all cases, while the pressure values has the maximum effect as the magnetic field angle is $\varphi = \pi/2$.  


\subsection{Limitations and Future Steps}
\begin{figure}[H]
    \centering
    \includegraphics[scale=0.25]{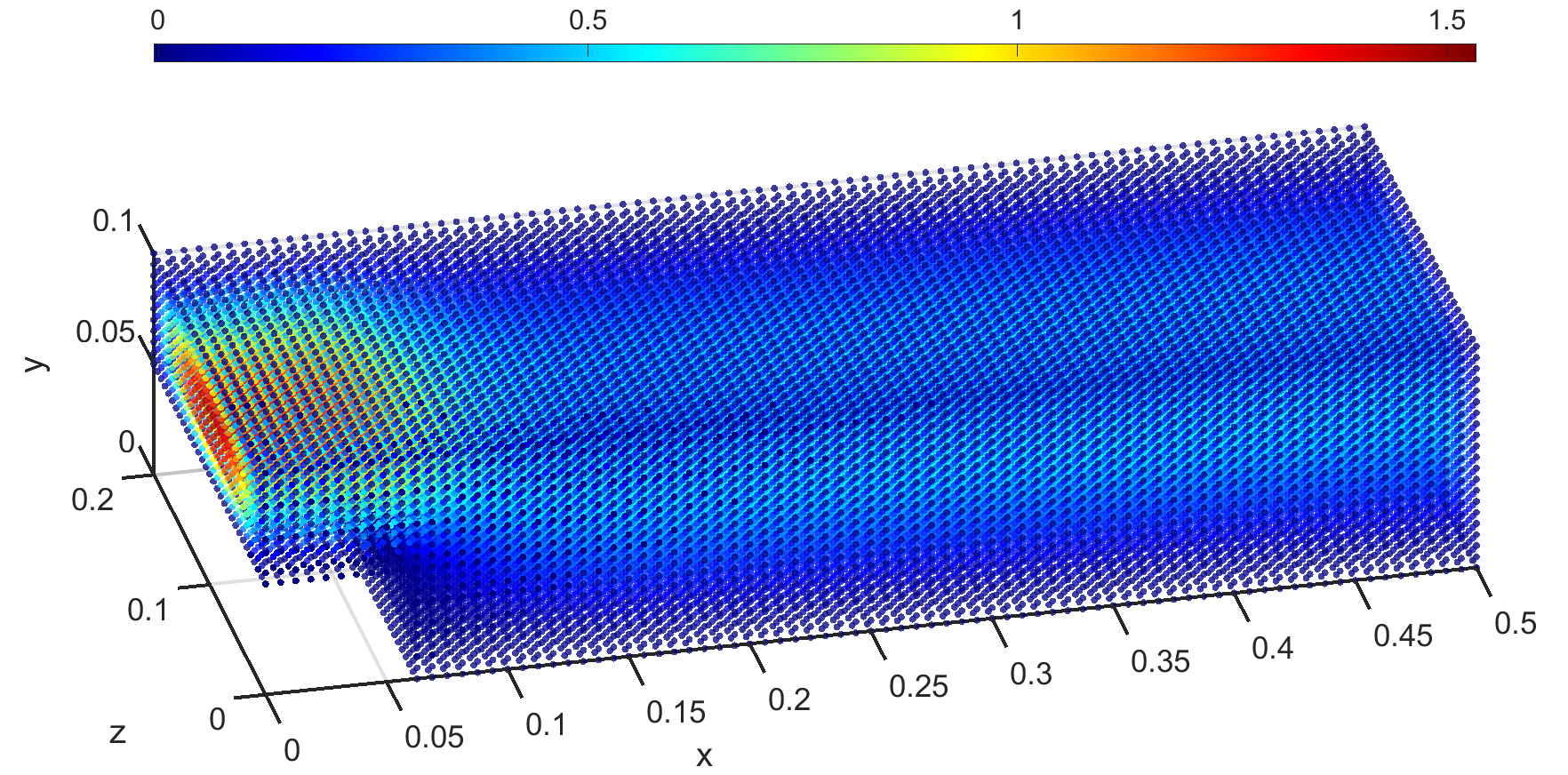}
    \caption{Scatter plot for the $u$-velocity component, in the case of the 3D Backward Facing step for $Re=100$.}
    \label{u3D}
\end{figure}

The study of BFS exhibits great interest in three dimensions due to the asymmetries observed in the reattachment points in the $z$-direction. The numerical procedure used can be extended to the three-dimensional case of BFS to compare with three-dimensional experimental studies. A preliminary study was conducted, and the boundary conditions are identical to the two-dimensional study, expanded in $z$-direction. The results depicted in Figure~\ref{u3D} show the $u$-velocity of the fluid. The results were produced by the same numerical algorithmic procedure, described in the present study, expanded in the three-dimensional case for a grid size of $79\times 29 \times 19$. 


Changing the attribute of the magnetic field from uniform to non-uniform would provide new insights into its effects in such a problem, since it is expected to locally disturb the vortex generated by the fluid flow and the geometry. Furthermore, an advanced turbulence model, utilizing two-equation models or Large Eddy Simulation, will be pursued to examine the robustness of the algorithm in situations that include extended recirculation regions with strongly nonlinear dynamics.

\section{Conclusions}

This work studies the accuracy and robustness of a numerical framework for simulating magnetohydrodynamic (MHD) flow over a BFS. The governing equations were discretized using a second-order FVM on a collocated grid and solved directly utilizing a Newton-like algorithm. The solver provides robustness and efficiency with reliable convergence. This approach is preferable to this problem, as intense source terms may result in a lack of convergence in the case of an inappropriate iterative approach. The numerical methodology was validated against a wide range of well-established hydrodynamic benchmarks, including reattachment lengths across several Reynolds numbers. The results showed very good agreement with previous numerical and experimental studies, confirming the accuracy and applicability of the proposed numerical approach.

The presence of the magnetic field reduced the fluid momentum and the reattachment length, and increased pressure due to the Lorentz force. Notably, the field orientation played a critical role, as the strongest effects were observed when the magnetic field was applied vertically $(\varphi = \pi/2)$, vanishing the secondary vortex that appears in the upper wall at higher Reynolds numbers. By systematically analyzing the impact of both magnetic field strength and direction, this study addresses a gap in the current literature on BFS/MHD flows. Furthermore, a preliminary extension to three-dimensional simulations demonstrates the robustness of the proposed solver and its potential for application in complex flow configurations and future studies involving turbulence or non-uniform magnetic fields.

\textbf{Acknowledgements:} This research was implemented in the framework of the Action “Flagship actions in interdisciplinary scientific fields with a special focus on the productive fabric”, which is implemented through the National Recovery and Resilience Fund Greece 2.0 and funded by the European Union–NextGenerationEU (Project ID: TAEDR-0535983).

\bibliography{references}  






\end{document}